\documentclass[preprint2]{aastex}

\newcommand{\vdag}{(v)^\dagger}

\slugcomment{Not to appear in Nonlearned J., 45.}

\shorttitle{Neutron star equations of state}
\shortauthors{Webb \& Barret}

\begin{document}


\title{Constraining the equation of state of supra-nuclear dense matter from XMM-Newton observations of neutron stars in globular clusters}


\author{Natalie A. Webb \altaffilmark{1} and Didier Barret\altaffilmark{1}}
\affil{Centre d'Etude Spatiale des Rayonnements, 9 Avenue du Colonel Roche, 31028 Toulouse Cedex 04, France}
\email{Natalie.Webb@cesr.fr}


\begin{abstract}
We report on the detailed modelling of the X-ray spectra of three
likely neutron stars.  The neutron stars, observed with XMM-Newton are
found in three quiescent X-ray binaries in the globular clusters:
\object{$\omega$ Cen}, \object{M 13} and \object{NGC 2808}.  Whether
they are accreting at very low rates or radiating energy from an
accretion heated core, their X-ray spectra are expected to be those of
a hydrogen atmosphere. We use and compare publicly available hydrogen
atmosphere models, with constant and varying surface gravities to
constrain the masses and radii of the neutron stars. Thanks to the
high XMM-Newton throughput, and the accurate distances available for
these clusters, using the latest science analysis software release and
calibration of the XMM-Newton EPIC cameras, we derive the most
stringent constraints on the masses and radii of the neutron stars
obtained to date from these systems. A comparison of the models
indicate that previously used hydrogen atmosphere models (assuming
constant surface gravity) tend to underestimate the mass and
overestimate the radius of neutron stars. Our data constrain the
allowed equations of state to those which concern normal nucleonic
matter and one possible strange quark matter model, thus constraining
radii to be from 8 km and masses up to 2.4 M$_\odot$.
\end{abstract}





\section{Introduction}

Forty years after their discovery, the nature of the material making
up the core of neutron stars remains largely unknown. At densities
above a few times the equilibrium density of nuclear matter, models
predict the existence of exotic components such as pion or kaon
condensates or unconfined quarks  \citep[e.g.][]{latt07}.  The
exciting idea that neutron stars may contain exotic forms of matter
makes them of prime interest, not only for astrophysics but for
physics in general.

Different equations of state of dense matter predict different maximum
masses and different mass-radius relationships. So far 
accurate mass measurements have been made for radio pulsars giving masses ranging from 1.18$\pm^{\scriptscriptstyle 0.03}_{\scriptscriptstyle 0.02}$ M$_\odot$, for the case of one of the pulsars in PSR J1756-2251 \citep{faul05} to 1.4414$\pm0.0002$ M$_\odot$, for PSR B1913+16 \citep{weis05}. Such low values can be accommodated by a wide
range of equations of states and do not provide a strong constraint
on the composition of dense matter. There is however growing evidence
that neutron stars, as massive as 2 M$_\odot$, may also exist, in
particular in accreting X-ray binaries, as inferred from the study of
the properties of kilo-Hz QPOs (Quasi-Periodic Oscillations) (e.g. the
neutron star in 4U 1636-536 which is estimated to have a mass of
1.9-2.1 M$_\odot$ \cite{barr05}) and binary millisecond pulsars that
are no longer accreting \citep[e.g. PSR J0751+1807, which has a mass
measured through relativistic orbital decay of 2.1$\pm$0.2 M$_\odot$,][]{nice05}. If the latter estimates are confirmed (by dynamical
mass estimates of the binary components), they would rule out some
exotic forms of matter, such as quarks or pion condensates.

While there are some accurate mass measurements already available, the
situation is different with radii which are much harder to
constrain. X-ray spectroscopy is one promising avenue to follow to
tackle this issue. So far, there has been one single (and still
debated) measurement of redshifted absorption lines in the combined
early and late type I X-ray burst spectra of the X-ray binary EXO
0748-676. The line energies are consistent with FeXXVI H$\alpha$ (n =
2-3) and FeXXV He$\alpha$  (n = 2-3), respectively, both at the same
redshift z = 0.350$\pm$0.005 \citep{cott02}. The
measured redshift provided a direct estimate of Mass/Radius (M/R)
($M/R=\frac{c^2}{2G}(1-(1+z)^{-2}$).  Assuming a canonical mass of 1.4
M$_\odot$, this would imply a radius of about 9 km. Following this
result, and using the burst properties, \cite{oeze06} estimated both
mass and radius separately. This led to a massive neutron star with a
mass M $\ge 2.10\pm 0.28$M$_\odot$ (and R$\ge 13.8\pm1.8$ km) in EXO
0748-676, a result suggesting again that if this system is typical,
exotic forms of matter, such as condensates and unconfined quarks do
not exist in neutron star cores.

Another promising way of inferring radii is through observations of
quiescent X-ray emission from neutron stars for which the distance (d) can
be estimated reliably ($F_\infty=(R_\infty/d)^2 \sigma T_\infty^4$,
where $F_\infty$, $T_\infty$ are the flux and radiation temperatures
redshifted to the Earth and $R_\infty=R/\sqrt{1-2GM/R c^2}$ is the
radiation radius). Whether the energy reservoir is the heat deposited
deep in the neutron star crust during the outburst phase of the
transient \citep{brow98}, or sustained by a low-level of radial
accretion \citep{vanp87} (via an advection dominated accretion flow), the X-rays originate from the atmosphere of
the neutron star. The spectrum radiated by the atmosphere will depend
on its actual composition, the strength and the structure of  the
magnetic field. In general, it is thought that the old neutron stars
in globular clusters have low magnetic fields and hydrogen-rich
atmospheres (gravitational settling of heavier elements occurs
rapidly, see \cite{rutl02} and references therein). Early non magnetic
hydrogen atmosphere models have been shown to provide adequate fits to
the quiescent X-ray spectra of several neutron star systems, all
providing plausible values for the neutron star radius
\citep[typically around 10 km,][]{rutl02,gend03a,gend03b}.  Compared
to neutron stars in the field, those in globular clusters have
accurate distance estimates, and the fitting of their very soft
spectra is made easier by the fact that at least the value of the
interstellar absorption derived from the optical extinction is well
known, even if the absorption intrinsic to the system is largely
unknown.

Following the work done by \cite{hein06} on Chandra data of the
quiescent neutron star X-ray binary X7 in \object{47 Tucanae}, we
apply recently improved neutron star atmosphere models, available in
the latest release of the XSPEC spectral fitting package
\citep[version 12.3.0,][]{arna96}, to three neutron stars in quiescent
X-ray binaries that we have observed in three globular clusters with
XMM-Newton. This is part of an on-going program to locate, identify
and classify faint X-ray sources in globular clusters \citep{serv07,gend03a,gend03b,webb04, webb06},
with an aim to determining the internal energy source that slows down
the inevitable collapse of globular clusters. Early fitting of the
X-ray spectra with the first hydrogen-atmosphere models of \cite{zavl96} assuming  a constant surface gravity (log g$_s$=14.385,
corresponding to a 1.4 M$_\odot$ and 10 km radius neutron star), were
reported for \object{$\omega$ Cen} and \object{M~13} in \cite{gend03a,gend03b}. However, as emphasised by \cite{hein06},
using models with appropriate surface gravity for each fitted value of
the mass and radius of the neutron star is important when interpreting
high quality X-ray spectra, which is indeed the case for the spectra
measured with XMM-Newton. Applying the improved models is the main
motivation of this paper, which further benefits from improved data
analysis software and calibration of the EPIC instruments compared to
the earlier published results on these sources.

In the next section, we describe the observations, recalling the
essential parameters of the clusters (distance, extinction,...), the
data reduction and present the results of the spectral fitting. We
use, for comparison purposes three different models, described
hereafter.

\section{Observations and data reduction}
\label{sec:obs}

We have observations of the three likely neutron stars in three
different globular clusters, \object{$\omega$ Centauri} ($\omega$
Cen), \object{M~13} and \object{NGC 2808}.  Observations of each of
these clusters were made with the X-ray observatory XMM-Newton.  All
three EPIC cameras were used in the full-frame mode with the medium
filter.  Further information about these observations can be found in
Table~\ref{tab:obs}.

\begin{table*}[!t]
  \caption[]{Summary of the globular clusters (GC) and their
  observations. }
\medskip
     \label{tab:obs}
       \begin{tabular}{lcccccc}
         \hline  \hline  \noalign{\smallskip}  GC & Date & Det. &
         T$_{obs}$ & GTI & Dist. & n$_H$\\  
\hline  
$\omega$ Cen & 2001 Aug 13 & M1 & 40 & 40 & 5.3 & 0.067 \\ 
& & M2 & 40 & 40 & & \\ 
& & pn & 40 & 40 & & \\ 
M 13 & 2002 Jan 28 & M1 & 17 & 12 & 7.7 & 0.011\\ 
 & & M2 & 17 & 11 & & \\ 
& & pn &  14 & 8 & & \\ 
M 13 & 2002 Jan 30 & M1 & 18 & 14 & & \\ 
& & M2 & 18  & 14 & & \\ 
& & pn & 13 & 8 & & \\ 
M 13 & 1990 Jun 1 & PSPC & 46 & 46 & & \\ 
NGC 2808 & 2005 Feb 2  & M1 &
         41 & 41 & 9.6 & 0.128\\ & & M2 & 41 & 41 & & \\ & & pn & 40 &
         31 & & \\ \hline
  \end{tabular}
\begin{minipage}{8cm}
\tablenotetext{}{Det. indicates the detector used (M1/M2= MOS 1 or 2),
  T$_{obs}$ gives the observation time in ks and GTI gives the good
  timing interval (in ks) after filtering for soft proton flares. The
  distances (Dist.)  are taken from \cite{harr99} and are given in
  kpc.  The column densities (n$_H$, $\times 10^{22}$ cm$^{-2}$) to
  these clusters are from \cite{harr99,pred95}.}
\end{minipage}
\vspace*{-0.3cm}
 \end{table*}

To reduce these data we used the latest version of the XMM-Newton
Science Analysis Software (SAS, version 7.0).  This has many
improvements over the earlier versions of the SAS used to reduce
\object{$\omega$ Cen} and \object{M~13} (version 5.3.3) such as
upgraded EPIC calibration, resulting in a much better cross
calibration among the EPIC instruments and which includes modelling of
spatial and temporal response dependencies. Improvements to the bad
pixel finding algorithms (decreasing the noise at low energies
significantly, important for neutron stars which emit mainly at low
energies), the  vignetting correction and exposure maps have also been
made (see the SAS release notes\footnote{
http://xmm.vilspa.esa.es/sas/7.0.0/documentation/releasenotes/}).  The
MOS data were reduced using the `emchain'.  The event lists were
filtered, so that 0-12 of the predefined patterns (single, double,
triple, and quadruple pixel events) were retained and the high
background periods were identified by defining a count rate threshold
above the low background rate and the periods of soft proton flares
were then flagged in the event list.  We also filtered in energy. We
used the energy range 0.2-10.0 keV, as recommended in the document
`EPIC Status of Calibration and Data Analysis' \citep{kirs02}.  The
{\it pn} data were reduced using the `epchain' of the SAS.  Again the
event lists were filtered, so that 0-4 of the predefined patterns
(single and double events) were retained, as these have the best
energy calibration.  We again filtered in energy, where we used the
energy range 0.2-10.0 keV and we also filtered for the soft proton
flares.  A summary of the observations and good time intervals for
each observation and camera is given in Table~\ref{tab:obs}.

\begin{figure}[!t]
\includegraphics[angle=0,scale=.40]{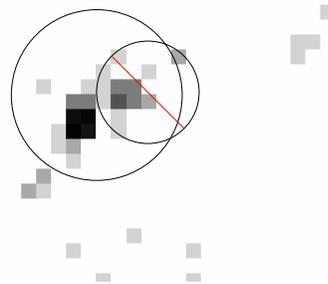}
\caption{Image showing the region used to extract the M 13 neutron star spectrum (brighter source) and the region excluded due to the source found at 18.4$\arcsec$
from the neutron star, superposed on the MOS 1 data (0.2-10.0 keV).}
\label{fig:M13ext}
\end{figure}

\begin{figure}[!t]
\includegraphics[angle=0,scale=.40]{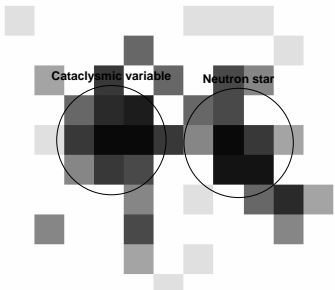}
\caption{MOS 1 image (0.2-10.0 keV) showing the region used to extract the NGC 2808 neutron star spectrum and it's proximity to the bright cataclysmic variable.}
\label{fig:NGC2808ext}
\end{figure}

We extracted the $\omega$ Cen spectra using circles of radii
$\sim$45\arcsec\ (to include at least 90\% of the available flux from
the neutron star) centred on the source.  For the neutron star in
\object{M 13} we used an extraction radius of $\sim$25\arcsec\ due its
close proximity to other sources and excluded a region of radius
15\arcsec\ around a second source that fell in the extraction
region (the large region contains at least 80\% of the total source
counts from the neutron star and less than 10\% of the counts from the
neighbouring source, \citet{ehle06}), see Fig~\ref{fig:M13ext}.  Due to over-crowding in the
centre of \object{NGC 2808} we used extraction regions of 8\arcsec,
which includes 50\% of the total source counts from the neutron star
\citep{ehle06}, see Fig~\ref{fig:NGC2808ext}.  We used a similar neighbouring region, free from
X-ray sources to extract a background spectrum.  We rebinned the MOS
data into 15 eV bins and the {\it pn}  data into 5eV bins as
recommended in \cite{ehle06}.  We used the SAS tasks `rmfgen' and
`arfgen' to generate a `redistribution matrix file' and an `ancillary
response file', for each spectrum.

\section{Spectral Analysis}
\label{sec:spec}

We exploit three different hydrogen atmosphere models that are
publicly available in Xspec 12.3.0.  These are the basic neutron star
atmosphere model (nsa) that we used previously in \cite{gend03a,gend03b}.  This model includes a uniform surface (effective)
temperature, either a non-magnetised neutron star or with a field B =
10$^{12}$ or B = 10$^{13}$ G and a radiative atmosphere in hydrostatic
equilibrium.  The nsagrav model used is similar, but allows for a
variety of surface gravitational accelerations, ranging from 10$^{13}$
to 10$^{15}$ cm s$^{-2}$, adapted to the masses and radii
investigated.  Finally we use the nsatmos model described by \cite{hein06}.  This model includes a range of surface gravities and
effective temperatures, and incorporates thermal electron conduction
and self-irradiation by photons from the compact object. It also
assumes negligible (less than 10$^9$ G) magnetic fields and a pure
hydrogen atmosphere.

For each neutron star, we fitted the spectrum obtained from all three
cameras simultaneously.  However, for the neutron star in \object{M
13} we also included the ROSAT PSPC archival observations of this
source in \object{M 13} \citep{verb01} as in \cite{gend03b}, as the
ROSAT data extends below the XMM-Newton data, down to 0.1 keV (only
11.5 ks of data exists in the archives for \object{$\omega$ Cen},
which is insufficient to improve our spectra.  No data exist for
\object{NGC 2808}). The ROSAT PSPC has a poorer angular resolution
than the EPIC cameras (PSPC = 25\arcsec\ and MOS = 6\arcsec) and the
spectrum extracted contains data from both the neutron star and the
source found at 18.4$\arcsec$ from the neutron star, see Fig.~\ref{fig:M13ext}.  We have
extracted the XMM-Newton EPIC MOS spectrum of this source.  The best
fit to this spectrum is either a power law,
$\Gamma$=1.75$\pm^{\scriptscriptstyle 0.61}_{\scriptscriptstyle
0.55}$, Cstatistic = 5.96 (8 bins) or a bremsstrahlung model,
kT=4.49$\pm^{\scriptscriptstyle 171.70}_{\scriptscriptstyle 2.60}$,
Cstatistic = 5.98 (8 bins).  This source contributes approximately
28\% of the counts in the ROSAT band (0.1-2.4 keV).  To take this
source into account, we have allowed the normalisation for the ROSAT
data to float, to account for the uncertain cross-calibration between
the two observatories.

\begin{table*}[!t]
\begin{minipage}{16cm}
\caption{Results of the spectral fitting for the three neutron stars
with the three different hydrogen atmosphere models, as described in
Section~\ref{sec:spec}.  }
\medskip
\label{tab:res}
\begin{tabular}{lcccc}
\hline \hline Cluster & NSA & NSAGRAV & NSATMOS & L$_{bol}$ (erg
s$^{-1}$)\\ \hline $\omega$ Cen & $N_H$=0.12$\pm^{\scriptscriptstyle
0.04}_{\scriptscriptstyle 0.02}$ & $N_H$=0.11$\pm^{\scriptscriptstyle
0.05}_{\scriptscriptstyle 0.03}$ &  $N_H$=0.12$\pm^{\scriptscriptstyle
0.04}_{\scriptscriptstyle 0.02}$ & 4.91 $\times$ 10$^{32}$\\ &
$log(T)$=5.99$\pm^{\scriptscriptstyle 0.30}_{\scriptscriptstyle 0.04}$
& $log(T)$=5.99$\pm^{\scriptscriptstyle 0.20}_{\scriptscriptstyle
0.11}$ &  $log(T)$=5.98$\pm^{\scriptscriptstyle
0.33}_{\scriptscriptstyle 0.10}$  & \\ &
$M$=1.76$\pm^{\scriptscriptstyle 0.74p}_{\scriptscriptstyle 1.26p}$ &
$M$=1.40$\pm^{\scriptscriptstyle 1.1p}_{\scriptscriptstyle 1.10p}$ &
$M$=1.66$\pm^{\scriptscriptstyle 0.84}_{\scriptscriptstyle 1.16p}$ & \\
& $R$=11.30$\pm^{\scriptscriptstyle 7.27}_{\scriptscriptstyle 6.30p}$ &
$R$=10.00$\pm^{\scriptscriptstyle 7.80}_{\scriptscriptstyle 4.00p}$ &
$R$=11.66$\pm^{\scriptscriptstyle 7.03}_{\scriptscriptstyle 4.99}$ &
\\ & Cstat= 91.28, 85 dof & Cstat= 91.26, 85 dof & Cstat= 91.35, 85
dof & \\ \hline M 13 & $N_H$=0.013$\pm^{\scriptscriptstyle
0.005}_{\scriptscriptstyle 0.005}$ &
$N_H$=0.013$\pm^{\scriptscriptstyle 0.005}_{\scriptscriptstyle 0.004}$
&  $N_H$=0.012$\pm^{\scriptscriptstyle 0.004}_{\scriptscriptstyle
0.003}$ & 5.08 $\times$ 10$^{32}$ \\ &
$log(T)$=6.00$\pm^{\scriptscriptstyle 0.01}_{\scriptscriptstyle 0.01}$
& $log(T)$=6.00$\pm^{\scriptscriptstyle 0.04}_{\scriptscriptstyle
0.02}$ & $log(T)$=6.00$\pm^{\scriptscriptstyle
0.01}_{\scriptscriptstyle 0.09}$ &  \\ &
$M$=1.38$\pm^{\scriptscriptstyle 0.08}_{\scriptscriptstyle 0.23}$ &
$M$=1.39$\pm^{\scriptscriptstyle 0.51}_{\scriptscriptstyle 0.67}$ &
$M$=1.30$\pm^{\scriptscriptstyle 0.06}_{\scriptscriptstyle 0.12}$ & \\
& $R$=9.95$\pm^{\scriptscriptstyle 0.24}_{\scriptscriptstyle 0.27}$ &
$R$=9.95$\pm^{\scriptscriptstyle 2.21}_{\scriptscriptstyle 0.36}$ &
$R$=9.77$\pm^{\scriptscriptstyle 0.09}_{\scriptscriptstyle 0.29}$ & \\
& $\chi^{\scriptscriptstyle 2}_{\scriptscriptstyle \nu}$=1.10, 62 dof
& $\chi^{\scriptscriptstyle 2}_{\scriptscriptstyle \nu}$=1.10, 62 dof
&  $\chi^{\scriptscriptstyle 2}_{\scriptscriptstyle \nu}$=1.08, 62 dof
& \\ \hline NGC 2808 &  $N_H$=0.17$\pm^{\scriptscriptstyle
0.05}_{\scriptscriptstyle 0.09}$ & $N_H$=0.18$\pm^{\scriptscriptstyle
0.11}_{\scriptscriptstyle 0.07}$ &  $N_H$=0.16$\pm^{\scriptscriptstyle
0.14}_{\scriptscriptstyle 0.05}$ & 1.02 $\times$ 10$^{33}$\\ &
$log(T)$=6.04$\pm^{\scriptscriptstyle 0.07}_{\scriptscriptstyle 0.14}$
& $log(T)$=6.08$\pm^{\scriptscriptstyle 0.06}_{\scriptscriptstyle
0.17}$ & $log(T)$=6.03$\pm^{\scriptscriptstyle
0.01}_{\scriptscriptstyle 0.25}$  & \\ &
$M$=0.67$\pm^{\scriptscriptstyle 0.59}_{\scriptscriptstyle 0.13}$ &
$M$=0.95$\pm^{\scriptscriptstyle 1.55p}_{\scriptscriptstyle 0.65p}$ &
$M$=0.91$\pm^{\scriptscriptstyle 1.60}_{\scriptscriptstyle 0.41p}$  &
\\ & $R$=8.45$\pm^{\scriptscriptstyle 0.36}_{\scriptscriptstyle 3.45p}$
& $R$=7.48$\pm^{\scriptscriptstyle 3.57}_{\scriptscriptstyle 1.48p}$ &
$R$=6.10$\pm^{\scriptscriptstyle 11.47}_{\scriptscriptstyle 1.10p}$  &
\\ & Cstat=17.59, 19 dof & Cstat=18.12, 19 dof & Cstat=16.95, 19 dof &
\\ \hline
\end{tabular}
\tablenotetext{}{ The column density, N$_H$ is $\times 10^{22}$
cm$^{-2}$, the temperature (T) is given as the logarithm of the
temperature in Kelvin, the masses are in solar units and the radii
in kilometres.  All errors are given at the 90\% confidence limit for
the one interesting parameter. To calculate the errors, the mass and distance were held fixed and all other parameters were allowed to vary, except when calculating the error on the mass, when the radius was held fixed.   A 'p' after the error value indicates that the hard limit of the model was reached.  The estimated unabsorbed bolometric
luminosity for each neutron star is also given.}
\end{minipage}
\end{table*}

\begin{figure}[!t]
\includegraphics[angle=0,scale=.4]{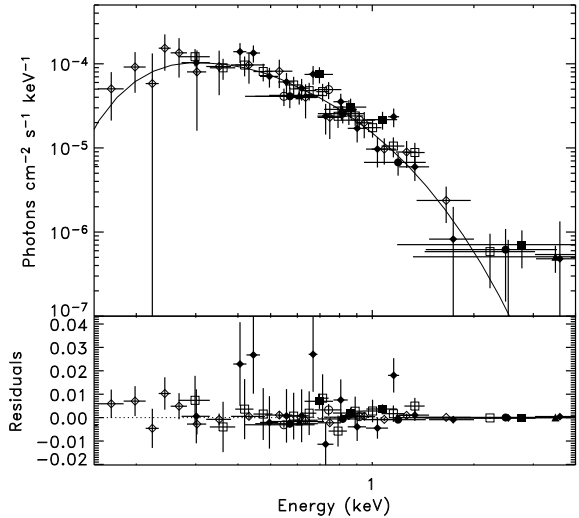}
\caption{Unfolded spectrum of the neutron star in M 13 with the best
nsatmos model fit and the residuals to the fit. The different symbols indicate the data from the different instruments used: open diamonds show PSPC data, filled circles and triangles show MOS 1 data, filled squares and empty circles show MOS 2 data and filled diamonds and empty squares show pn data.}
\label{fig:m13}
\end{figure}

All the data were binned to contain a minimum of 20 counts per bin
where possible and we used the $\chi^2$ technique to judge the
goodness of the model fits.  However, for \object{$\omega$ Cen} and
\object{NGC 2808} the data were binned with 15 counts per bin (to
increase the number of data points).  In these cases we used the
C-statistic to judge the goodness of the fit, as the number of counts
per bin is no longer strictly in the Gaussian statistics regime but
approaches that of the Poissonian statistics regime. Even though we
are at the limit between the two regimes, the C-statistic has been
shown to work well at even higher counts \citep{nous89} and should
therefore give good results.  The inverse is not necessarily
true. \cite{nous89} showed that using the Levenberg-Marquardt
algorithm (the algorithm used in Xspec) for small numbers of events per bin gives a systematic bias.

\subsection{Absorption in the spectra}

For each
model we included (photoelectric) absorption along the line of sight
to the object. This was initially fixed at the value determined for
each cluster (see Table~\ref{tab:obs}) and then allowed to vary.  In
every case we found that we required additional absorbing material,
assumed to be additional gas intrinsic to the system.  This was
typically 30-60\% more i.e. $\omega$ Cen required an n$_H$$\sim$0.11
$\times 10^{22}$ cm$^{-2}$, an increase of 60\% with a ftest
probability of 0.0099, indicating that it is reasonable to include
additional material. The normalisation was fixed to the value
corresponding to the distance to the cluster (see Table~\ref{tab:obs})
and the masses and radii were initially frozen to 1.4 M$_\odot$ and
10.0 km, but then allowed to vary.

We then tried to constrain the nature of the additional absorbing
material, in the same way as Heinke et al. (2006), by employing the
{\it Xspec} model {\em vphabs} and using the abundances corresponding to those
of each of the clusters.  For \object{M 13} we chose the iron abundance to be
3\% solar ([Fe/H] = -1.5), the abundances of C, N and O to be 5\%
solar ([X/H] = -1.32), the abundances of Ne to Ca to be 4\% solar
([X/H] = -1.38) and that of helium to be of solar abundance, following
Cohen \& Melendez (2005).  For $\omega$ Cen the situation is more
complicated as there are three distinct populations of stars in this
cluster, see e.g. Origlia et al. (2003).  We elected the abundances
from the largest population in the cluster, basing our choice on the
fact that the highest probability was that the donor star comes from
the largest population.  We adopt the iron abundance to be 3\% solar
([Fe/H] = -1.58), the abundances of C, N and O to be 6\% solar ([X/H]
= -1.24), the abundances of Ne to Ca to be 6\% solar ([X/H] = -1.19)
and again that of helium to be of solar abundance, following Origlia
et al. (2003) and Norris (2004).  For NGC 2808 we select the iron
abundance to be 9\% solar ([Fe/H] = -1.06), the abundances of C, N and
O to be 32\% solar ([X/H] = -0.5), the abundances of Ne to Ca to be
16\% solar ([X/H] = -0.8) and that of helium to be of solar abundance,
following Castellani et al. (2006) and Gratton (1982).  Again, however, it is
believed that NGC 2808 has as many as three populations of stars, with
different abundances (Piotto et al. 2007).  We chose the abundances for the only population with sufficient information in the literature.
We then modelled the neutron star spectra fixing the photoelectric absorption
(phabs) to be that of the cluster and allowing the
photoelectric absorption with variable abundances (vphabs) to vary (using the abundances given
above).  The results that we obtained gave very small masses and/or
radii, often small enough to hit the lower mass or radius limit
allowed in the models (see below for values).  

\begin{figure}[!t]
\includegraphics[angle=0,scale=.4]{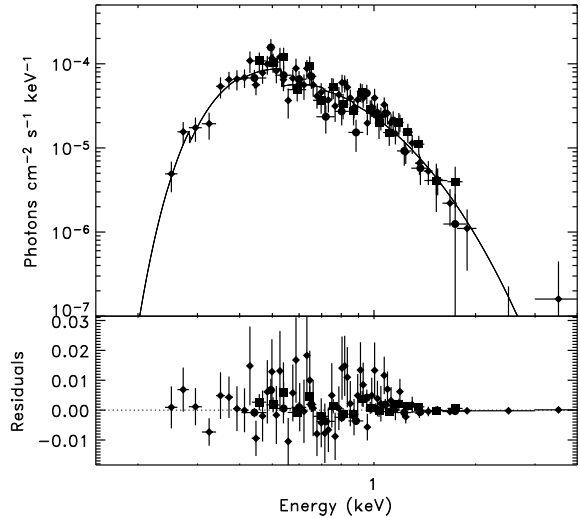}
\caption{Unfolded spectrum of the neutron star in $\omega$ Cen with
the best nsatmos model fit and the residuals to the fit.  The different symbols indicate the data from the different instruments used: filled circles show MOS 1 data, filled squares show MOS 2 data and filled diamonds show pn data.}
\label{fig:ocen}
\end{figure}

\begin{figure}[!h]
\includegraphics[angle=0,scale=.4]{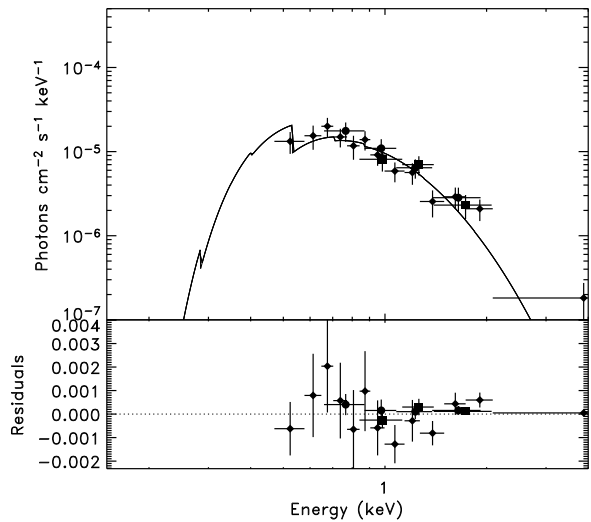}
\caption{Unfolded spectrum of the neutron star in NGC 2808 with the
best nsatmos model fit and the residuals to the fit. The different symbols used are the same as those employed in Fig.~\ref{fig:ocen}.}
\label{fig:ngc2808}
\end{figure}

We thus decided to
determine whether our data are of sufficiently good quality to
constrain the nature of the additional absorbing material. We
simulated a spectrum similar to that obtained for $\omega$ Cen, using the simple phabs model.  We modelled
this spectrum in {\em Xspec} using first the phabs model and then the vphabs
model.  We obtained similar $\chi^{\scriptscriptstyle
2}_{\scriptscriptstyle \nu}$ values (1.17, 78 dof when using phabs and
1.29, 78 dof when using vphabs), but again the masses and radii were
very low when the vphabs model was employed, only $\sim$80\% of the
value given to simulate the spectrum as opposed to values within 10\%
when using the phabs model.  We also simulated the same spectrum but using the vphabs model.  Again, modelling
this spectrum in Xspec using the vphabs model we found values as low as 33\% of the original value, yet when modelling with the phabs model our values are
within 5\% of our original values.  Further, we obtain equally good
fits using vphabs to our simulated spectrum ($\chi^{\scriptscriptstyle
2}_{\scriptscriptstyle \nu}$=1.09, 78 dof) when we choose abundances
that are different by 300\% to those used in the simulated spectrum,
and we still have the problem that either the mass or the radius is
very small.  We therefore conclude that the quality of our data is
insufficient to constrain the nature of the additional absorbing
material and we use the phabs model only.

\subsection{Contribution from neighbouring sources}

We also investigated whether we required an extra power law tail,
especially for the two neutron stars found in proximity of other
sources i.e. the neutron stars in M 13 and NGC 2808.  This is
important as substantial flux from other sources is likely to be the
dominant source of counts in the higher energy bands (i.e. above 1
keV).  This may tend to systematically bias the spectral fits to
higher temperatures, and thus smaller radii.

To do this we extracted the spectra of the neighbouring sources and
determined their photon indices ($\Gamma$$\sim$1.75 for M 13 and
$\Gamma$$\sim$1.5 for NGC 2808).  We then estimated the contribution from the neighbouring sources 
in the extraction region for each neutron star (8\% for M 13 and 20\%
for NGC 2808, see Figs~\ref{fig:M13ext} and \ref{fig:NGC2808ext}).  We used these values and refitted our
spectra.  Interestingly, we found values very similar to those
obtained when no power law was included.  Even allowing the
normalisation to increase, increasing the weight of the power law,
made little difference to the fits.  Ftests show that adding the power
law is not necessary.  Probability results for fitting nsa, nsatmos and
nsagrav models with and without the additional power laws to the neutron star in M 13 are
0.31, 0.22 and 0.199 respectively and to the neutron star in NGC 2808 are 0.13, 0.06 and
0.08 respectively.  We therefore conclude that adding a
power law is not required by the data.

The spectra of the three neutron stars can be found
in Figs.~\ref{fig:m13}-\ref{fig:ngc2808}, along with the best fitting
nsatmos model.

\subsection{Modelling the spectra}

The principal goal of our spectral fitting is to self-consistently
constrain the allowed space in mass and radius using our three neutron
stars, but also to test the reliability and accuracy of the three
models examined.  Table~\ref{tab:res} gives the best fits for each
neutron star fitted with each of the three models, along with the best
fitting values for the column densities, surface temperatures of the
neutron stars, masses and radii. Errors are also given at the 90\%
confidence limit for the one interesting parameter. To calculate the errors, the mass and distance were held fixed and all other parameters were allowed to vary, except when calculating the error on the mass, when the radius was held fixed instead of the mass.  We also give the
estimated bolometric luminosity of the neutron star.  We found that we
did not require any additional parameters, such as lines or edges to
fit these data. However, this may be due to the quality of the data,
where deeper observations may reveal spectral features that will be
useful to constrain the gravitational redshift at the NS surface
\citep{brow98,rutl02,hein03}.

\begin{figure}[!ht]
\hspace*{-0.5cm}\includegraphics[angle=0,scale=.43]{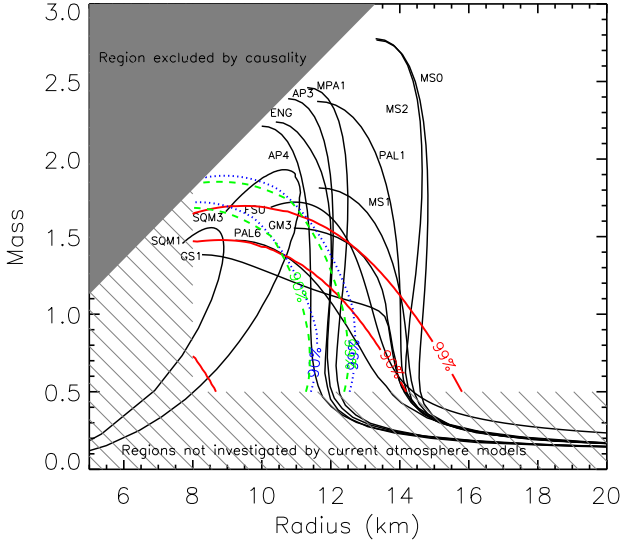}
\caption{Contour plot showing the results of modelling the neutron
star in M 13 with the xspec models: nsa (solid/red line), nsatmos
(dotted/blue line) and the nsagrav (dashed/green line). The 90\% and
99\% confidence contours for the neutron star masses (M$_\odot$) and
the radii (km) are plotted along with neutron star equations of state
taken from \cite{latt07} }
\label{fig:M13cont}
\end{figure}

\begin{figure}[!h]
\hspace*{-0.5cm}\includegraphics[angle=0,scale=.43]{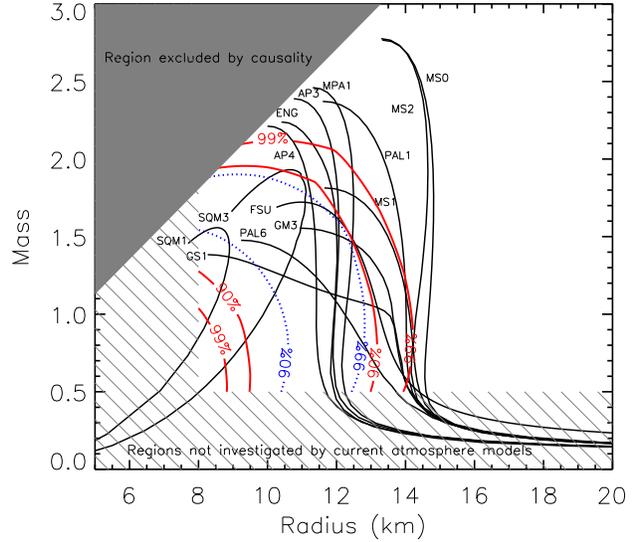}
\caption{Contour plot showing the results of modelling the neutron
stars in $\omega$ Cen (solid/red line)  and NGC 2808 (dotted/blue
line) with the xspec model nsatmos. The 90\% and 99\% confidence
contours for the neutron star masses (M$_\odot$) and the radii (km)
are plotted along with neutron star equations of state taken from
\cite{latt07} }
\label{fig:ocencont}
\end{figure}

\begin{figure}[!h]
\hspace*{-0.5cm}\includegraphics[angle=0,scale=.43]{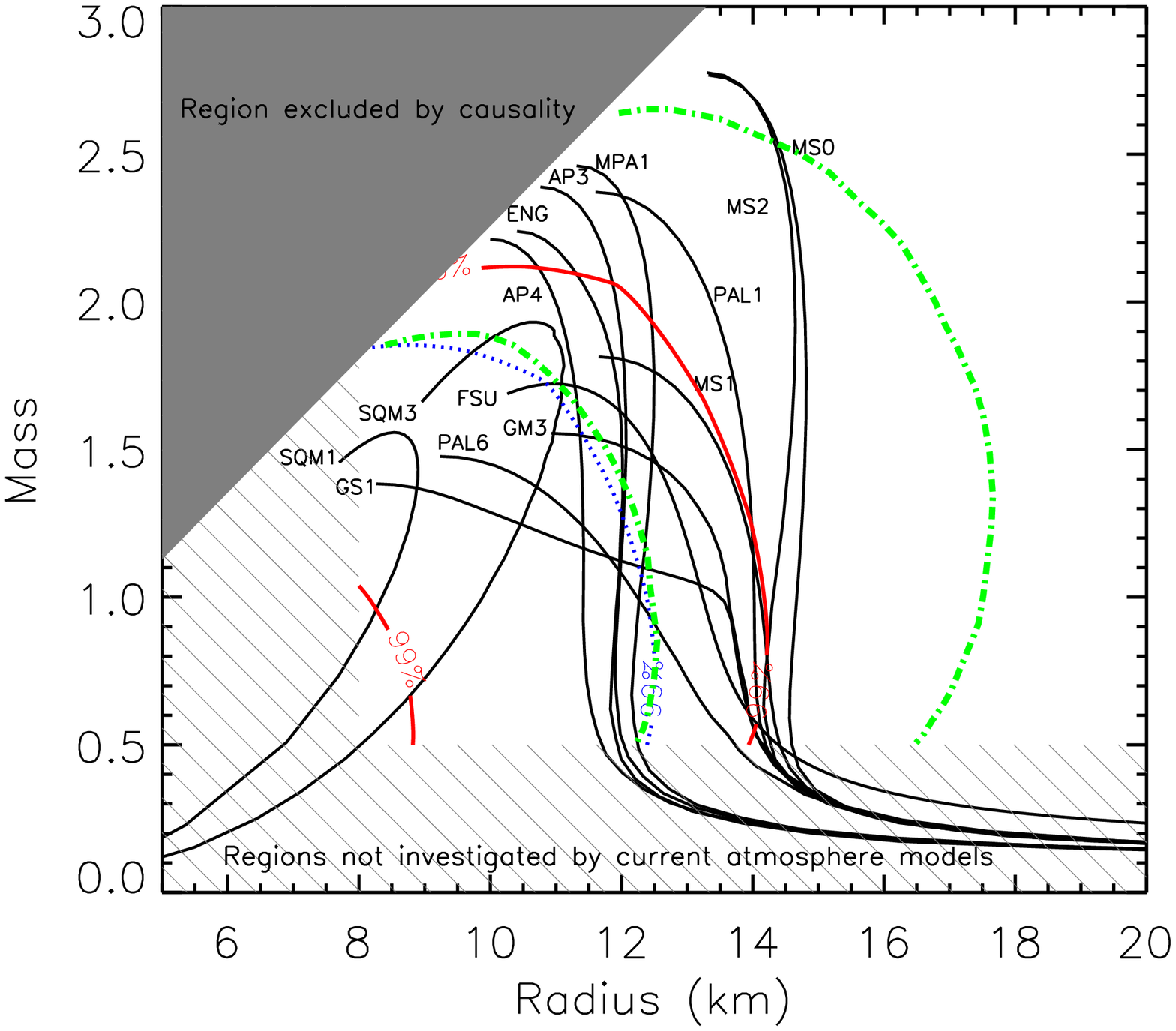}
\caption{Contour plot showing the results of modelling the neutron
stars in M 13 (dotted/blue line), $\omega$ Cen (solid/red line)  and
X7 in 47 Tuc (dashed-dotted/green line) \citep{hein06} with the xspec
model nsatmos. The 99\% confidence contours for the neutron star
masses (M$_\odot$) and the radii (km) are plotted  along with neutron
star equations of state taken from \cite{latt07}. The grey hashed
region indicates the region not investigated with the models.}
\label{fig:3cont}
\end{figure}

We use the steppar command in XSPEC to vary both the radius and mass
parameters simultaneously, allowing the temperature to vary as well to
find the best fit. The minimum (and maximum) radii allowed with these
models are 5 km (30/20 km) for the nsatmos and nsa models respectively
and 6 km (20 km) for the nsagrav model.  We chose an inferior limit of
8 km and a superior limit of 18 km as all three models gave stable
results when fitting the data in these regions.  The minimum (and
maximum) masses allowed with these models are 0.5M$_\odot$
(3.0/2.5M$_\odot$) for the nsatmos  and nsa models respectively and
0.3M$_\odot$ (2.5M$_\odot$) for the nsagrav model.  We chose the
region between 0.5 and 2.3M$_\odot$ as again all three models gave
stable results in these regions.  We show  90\% confidence, and 99\%
confidence ($\Delta \chi^2$ = 4.61, and 9.21 respectively) contours in
neutron star mass and radius.  Fig.~\ref{fig:M13cont} shows the
results of modelling the neutron star in \object{M 13} with all three
neutron star atmosphere models.  This figure indicates that the two
models, nsatmos and nsagrav, give comparable results, whereas the nsa
model gives differing results.  The nsa model is more accurate in
constraining R$_\infty$ than the nsatmos and nsagrav models, due to a
degeneracy in spectral shape variations between the surface gravity
and the surface temperature.  However, here we calculate a range of
neutron star masses and radii, for which the nsatmos and nsagrav,
thanks to their variable surface gravities, are better adapted \citep[see
the discussion in][]{hein06}.  The use of these fixed gravity
models for testing neutron stars with a variety of masses and radii
(and therefore gravities) is not therefore strictly appropriate.  This
is borne out by the modelling. For this reason and to simplify the
following plots, we ignore the nsa model (as it is less accurate) and
we plot only the nsatmos model (as the nsagrav and nsatmos models are
similar) to show the constraints on mass and radius determined by
modelling \object{$\omega$ Cen} and \object{NGC 2808}
(Fig.~\ref{fig:ocencont}).  The loci of models for the equations of
state for dense matter are those described in \cite{latt07} and
\cite{latt01} which include diverse equations  such as: SQM -
\cite{prak95}, a Strange Quark Matter model, PAL - \cite{prak88} a
neutron and proton model using a schematic potential, GM -
\cite{glen91}, a model containing neutrons, protons and hyperons using
a field theoretical approach, GS - \cite{glen99}, a model containing
neutrons, protons and kaons using a field theoretical approach.

\section{Discussion and Conclusions}
\label{sec:discuss}

As seen in Section~\ref{sec:spec} even if the nsa model is more accurate in constraining R$_\infty$ than the nsatmos and nsagrav models, when calculating a range of neutron star masses and radii, we find that the nsatmos and nsagrav models  are better adapted.  We
conclude that the use of fixed gravity models for testing neutron
stars with a variety of masses and radii (and therefore gravities) is
not strictly appropriate as previously indicated by \cite{gaen02} and
\cite{hein06}.

Using the results from Figs.~\ref{fig:M13cont} and \ref{fig:ocencont},
and the results found by \cite{hein06} when fitting Chandra
observations of X7 in 47 Tuc, we show the allowed equations of state
in Fig.~\ref{fig:3cont}.  Modelling the neutron star in \object{M 13}
alone shows that the data do not favour the stiffer equations of
state, such as MS0-2 and PAL 1.  We combine these results with those
from modelling the neutron star in \object{$\omega$ Cen}. The low
quality  data for the neutron star in \object{NGC 2808} results in
very poor constraints for this object and hence we disregard this
source in the discussion, although the results seem to be similar to
those for the neutron star in \object{M 13}.  The equations of state
that are satisfied by all three neutron stars fall in the middle of
the diagram and includes the equations of state of normal nucleonic
matter and one strange quark matter model.  The equations allowed are
GS1, PAL 6, AP3 \& 4, GM3, FSU, SQM3, ENG and MPA1, with radii above 8
km and masses up to 2.4 M$_\odot$.

\acknowledgments

This article was based on observations obtained with XMM-Newton, an
ESA science mission with instruments and contributions directly funded
by ESA Member States and NASA.  The authors thank J. Lattimer for providing them with the mass-radius relationships for the different equations of state shown in Figures~\ref{fig:M13cont}-\ref{fig:3cont} and B. Gendre for help
with the ROSAT spectrum.  The authors also acknowledge the CNES for
its support in this research and thank the (anonymous) referee for very valuable comments that have enabled us to improve the quality of this paper.

\end{document}